\documentclass[aps,superscriptaddress,pra,color]{revtex4-2}

\usepackage[english]{babel}
\usepackage[pdftex, pdftitle={Article}, pdfauthor={Author}]{hyperref} 

\usepackage{physics}
\usepackage[dvipsnames]{xcolor}
\hypersetup{colorlinks,linkcolor={blue!50!black},citecolor={blue!80!black}, urlcolor={blue!50!black}}
\usepackage{graphicx}
\usepackage{amsmath, amssymb, amsfonts}
\usepackage{bm}
\usepackage[letterpaper,margin=1in]{geometry} 
\usepackage[utf8]{inputenc}
\usepackage[section]{placeins}
\usepackage{float}
\usepackage{siunitx}
\newcommand{\mycomment}[1]{}
\usepackage[nameinlink]{cleveref}

\graphicspath{{Figures/}}

\begin{document}

\title{Helical locomotion in dilute suspensions}

\author{Albane Th\'ery}
\altaffiliation{These authors contributed equally}
\affiliation{Department of Mathematics, University of Pennsylvania, Philadelphia, Pennsylvania, USA}
\affiliation{Department of Applied Mathematics and Theoretical Physics, University of Cambridge,  Cambridge CB3 0WA, UK}

\author{Andres Zambrano}
\altaffiliation{These authors contributed equally}
\affiliation{School of Engineering, Brown University, Providence, Rhode Island, USA}

\author{Eric Lauga}
\email[]{e.lauga@damtp.cam.ac.uk}
\affiliation{Department of Applied Mathematics and Theoretical Physics, University of Cambridge,  Cambridge CB3 0WA, UK}

\author{Roberto Zenit}
\email[]{zenit@brown.edu}
\affiliation{School of Engineering, Brown University, Providence, Rhode Island, USA}

\date{\today}

\begin{abstract}
Motivated by the aim of understanding the effect of media heterogeneity on the swimming dynamics of flagellated bacteria, we study the rotation and swimming of rigid helices in dilute suspensions experimentally and theoretically. 
We first measure the torque experienced by, and thrust force generated by,  helices rotating without translating in suspensions of neutrally buoyant particles with varying concentrations and sizes. 
Using the ratio of thrust to drag forces $\xi$ as an empirical proxy for propulsion efficiency, our experiments indicate that $\xi$ increases with the concentration of particles in the fluid, with the enhancement depending strongly on the geometric parameters of the helix. 
To rationalize these experimental results, we then develop a dilute theoretical approach that accounts for the additional hydrodynamic stress generated by freely suspended spheres around the helical tail. 
We predict similar enhancements in the drag coefficient ratio and propulsion at a given angular speed in a suspension and study its dependence on the helix geometry and the spatial distribution of the suspended spheres.
These results are further reinforced by experiments on freely swimming artificial swimmers, which propel faster in dilute suspensions, with speed increases over $60 \%$ for optimal geometries. 
Our findings quantify how biological swimmers might benefit from the presence of suspended particles, and could inform the design of artificial self-propelled devices for biomedical applications. 

\end{abstract}


\maketitle

\section{Introduction} \label{sec:intro}

Motility allows organisms to explore new environments, search for food, reproduce, or escape competition~\cite{Bray2000, Sibona2007}, and efficient propulsion can be a significant evolutionary advantage~\cite{Wadhwa2021}. Many unicellular and multicellular microorganisms have the ability to swim, and an increasing number of propulsion modes at small scales have been described as more bacteria and plankton species are observed and quantified~\cite{Guasto2012, Miyata2020}. Understanding their locomotion is essential for biomedical and ecological applications, including preventing bacterial infections~\cite{Mathijssen2019}, understanding fertility~\cite{Dcunha2020}, or modeling interactions between microorganism populations in the ocean or soil~\cite{Wisnoski2023}. 
Conversely, robust predictions of viscous propulsion efficiency are necessary to design and control artificial or hybrid microswimmers~\cite{Tsang2020, Pauer2021} for micro-robotics and biomedical applications, such as drug delivery~\cite{Park2017}, imaging~\cite{Min2019} or microsurgery~\cite{Bunea2020}. 

The fundamental study of biological microscale locomotion has been extremely successful, drawing tools from mathematical fluid mechanics applied to simple, Newtonian fluids~\cite{lauga2009, Du2012,lauga2020}. However, microorganisms inhabit diverse and heterogeneous environments, such as soil~\cite{theves2015} and tissues~\cite{heddergott2012}, and they may self-propel in fluids rich in suspended polymers and particles, such as blood~\cite{heddergott2012} or mucus~\cite{ottemann1997, denissenko2012}. 
These complex fluids significantly deviate from the idealized bulk fluids considered in classical experiments and theory, and exhibit combinations of non-Newtonian behaviors, such as shear-dependent viscosity and viscoelasticity~\cite{Fauci2006, roselli2011, Arratia2022, Spagnolie2023}. 
The non-Newtonian properties of the surrounding fluid are known to deeply affect microscale propulsion~\cite{Spagnolie2015, Arratia2022, Spagnolie2023}, with past work showing that, depending on the propulsion mode and fluid properties, they can lead to an increase~\cite{Li2015, Gomez2016, Demir2020, kamdar2022a} or a decrease~\cite{shen2011, Datt2015} in swimming speed.

The presence of suspended obstacles has been proposed as a key factor explaining enhanced swimming speeds in suspensions, but also in polymeric fluids~\cite{kamdar2022a,kamdar2022}.
In suspensions, the swimming speed of micro-organisms has been experimentally reported to increase monotonically with particle concentration~\cite{jung2010, Juarez2010,patteson2015,park2016}. However, the mechanism for this increase has not been systematically characterized.
Some studies have been conducted in colloidal suspensions, in which random thermal fluctuations and inter-particle interactions are not negligible. This introduces a shear-rate dependence on viscosity~\cite{pan2010,hoffman1972} and significant particle cluster migration~\cite{shaqfeh2019,semwogerere2007,hoffman1972}, causing the fluid to exhibit simultaneously shear-rate-dependent viscosity, flow instability, and heterogeneity, with respective effects hard to identify and isolate. 
Additionally, while the speed enhancement of a given species of biological microswimmers can stem from general physical properties that apply to a wide range of biological and artificial swimmers, such as a higher efficiency of helical or wave-propagation propulsion, it can also be an organism-specific property, as for the proposed realignment of bacterial flagella and head for \textit{E. coli} in colloidal fluids~\cite{kamdar2022a}, and therefore less widely applicable.

From a theoretical standpoint, understanding swimmers in non-Newtonian fluids~\cite{fu2007,lauga2009,riley2014}, and suspensions in particular, poses major challenges and various modeling efforts have been proposed. 
The first approach has been to model the fluid as an effective continuous medium, leading to predictions that can depend on the choice of the empirical model for the surrounding fluid~\cite{Spagnolie2023}. The locomotion is different if the suspension is modeled, for example, as a fluid with shear-dependent viscosity~\cite{Hewitt2017}, a two-fluid medium~\cite{Du2012, Narayanan2024}, or, more frequently, as a porous or medium a gel with an effective Brinkman equation~\cite{leshansky2009, Leiderman2016, nganguia2018, Chen2020}. 
Alternatively, fluid-mediated interactions with obstacles can be modeled individually, but previous work considering detailed hydrodynamic interactions with individual particles has focused on arrays of stationary obstacles as opposed to suspended ones~\cite{leshansky2009}.

To isolate the physical impact of suspended particles on microscale locomotion, it is essential to conduct systematic studies with well-characterized swimmers and controlled fluid suspensions that do not exhibit shear-rate-dependent viscosity. We also need new models that accurately reproduce the physics of the non-Newtonian fluids and account for the hydrodynamic interactions with freely suspended objects rather than continuous approximations. 

In this paper, we propose a controlled setting for studying the propulsion of rigid helices in a non-Brownian suspension of rigid particles, including both experiments (with two distinct setups: helix with constrained motion and free swimmer) and theoretical (analytical and numerical) models for the interaction of helices with freely suspended particles. In section~\ref{sec:experiment}, we first measure both drag and thrust forces on a helix rotating without translation in the presence of suspended rigid spheres in a viscous background fluid~\cite{boyer2011}. 
We then quantify the propulsion efficiency by introducing a drag coefficient ratio that compares thrust and drag forces. 
We reveal a systematic increase in helical propulsion efficiency for dilute suspensions, with a strong dependence on the geometry of the helix. 
This effect is next studied theoretically and numerically in section~\ref{sec:theory}. We develop an analytical modification to resistive-force theory in the presence of suspended spheres for an infinite cylinder moving through the suspension,  accounting mathematically for the fact that the particles are free to move rather than fixed in an external matrix. 
To include local effects and geometry, we then extend our model to numerical simulations of a slender filament in a suspension of hard spheres and make quantitative predictions matching experiments. 
We finally test our experimental and theoretical predictions for free-swimming artificial swimmers actuated by a rotating magnetic field~\cite{godinez2012} in section~\ref{sec:swimmers}. We measure a significant increase in propulsion speed, consistently with our modified model for a free swimmer.

\section{Forces and torques on a rotating helix} \label{sec:experiment}

In this section, we outline our experimental setup to measure the propulsive efficiency of helical swimming in a suspension.
We measure the force and torque on a rigid helical flagellum held fixed in a rotating tank filled with a viscous fluid, with or without suspended particles.

\subsection{Experimental setup}

We take direct measurements of the forces and torques experienced by a helix as the surrounding fluids with suspended particles rotate at a fixed angular velocity. We first described the details of the apparatus, the parameter space explored, and the direct force and torque measurements in section~\ref{subsec:forceexp}. The viscous fluids and suspensions used in the experiments and their rheological properties are then discussed in section~\ref{subsec:susp_rheo}.

\subsubsection{Force and torque measurements} \label{subsec:forceexp}
\begin{figure}[t]
        \centering
\includegraphics[width = 0.8 \textwidth]{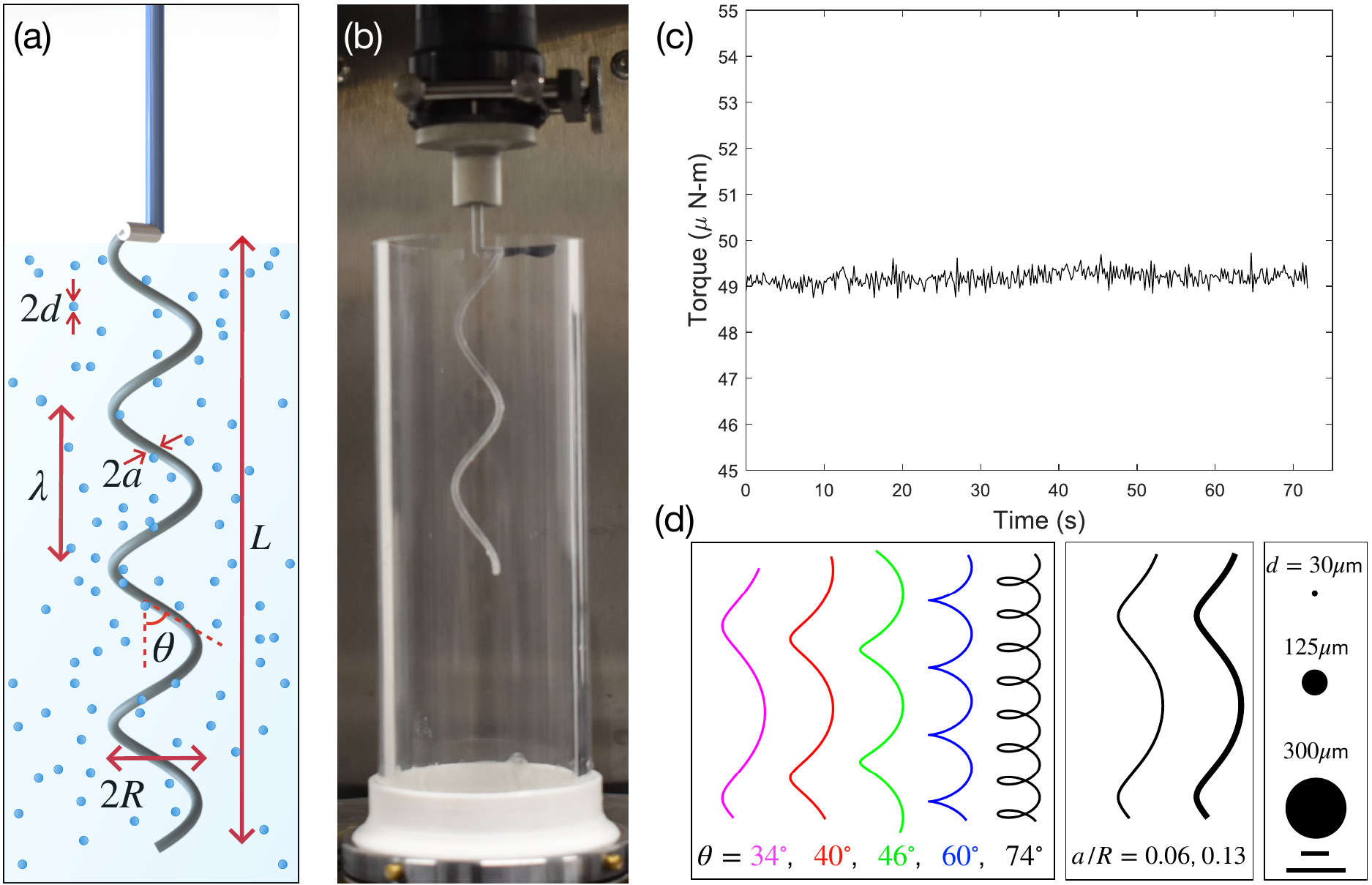}
     \caption{Experimental setup. (a) Schematic of helix geometry, with connecting rod at the top used to interface with the rheometer. The helix is characterized by its length $L$, radius $R$, thickness $2a$ and pitch angle $\theta$, with a wavelength $\lambda = 2 \pi R / \tan \theta$. 
     (b) Picture of the experimental setup with the tank, which rotates at a controllable angular velocity.
(c) Sample data collected from the experiment. 
(d) Parameter space explored in the experiments: we vary the helix pitch angle $\theta$, thickness $ 2a$, and the particle radius $d$. }
        \label{fig:force_setup}
    \end{figure}

In all experiments, the helix is held stationary while the tank rotates, with no translation of the helix. The helix geometry is shown schematically in Fig.~\ref{fig:force_setup}a. The radius $R = 4.5 \, \mathrm{mm}$ and length $L=65 \, \mathrm{mm}$ are constant, and we vary the pitch angle by modifying the pitch length $\lambda$. Given that $\tan\theta = 2 \pi R / \lambda $, the set of pitch angles tested in our experiments is $\theta= \{34^\circ,40^\circ,46^\circ,60^\circ,74^\circ\}$. We also vary the filament radius $a$ so that $a/R = \{0.06, \; 0.13 \}$. 

The helix is mounted on a rheometer (TA ARES G2) which serves as a force and torque transducer to collect measurements. The setup and some sample data collected from the instrument are shown in Fig.~\ref{fig:force_setup}b-c. 
The angular velocity is set at $\Omega = 4 \, \mathrm{s}^{-1}$, and all experiments are conducted at low Reynolds numbers $Re = \rho \Omega R^2 / \mu < 10^{-1}$, with $\rho$ and $\mu$ the density and dynamic viscosity of surrounding fluid respectively. The (volume) concentration of suspended particles $\phi=V_{\mathrm{solid}} / V_{\mathrm{total}}$ is varied in intervals of 5\% and kept dilute, namely $\phi = \left\{0,0.05,0.10,0.15,0.20 \right\}$. We test particles with three different radii $d$: $30 \, \si{\mu m}$, $125 \, \si{\mu m}$, and $300 \, \si{\mu m}$.

\subsubsection{ Rheological characterization of the suspensions} \label{subsec:susp_rheo}

\begin{figure}[t]
    \centering
    \includegraphics[width=0.32\textwidth]{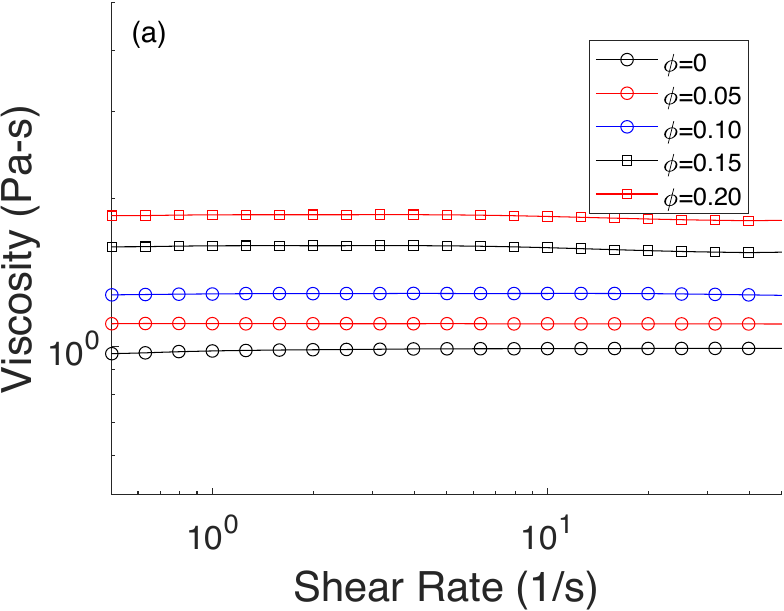}
    \includegraphics[width=0.32\textwidth]{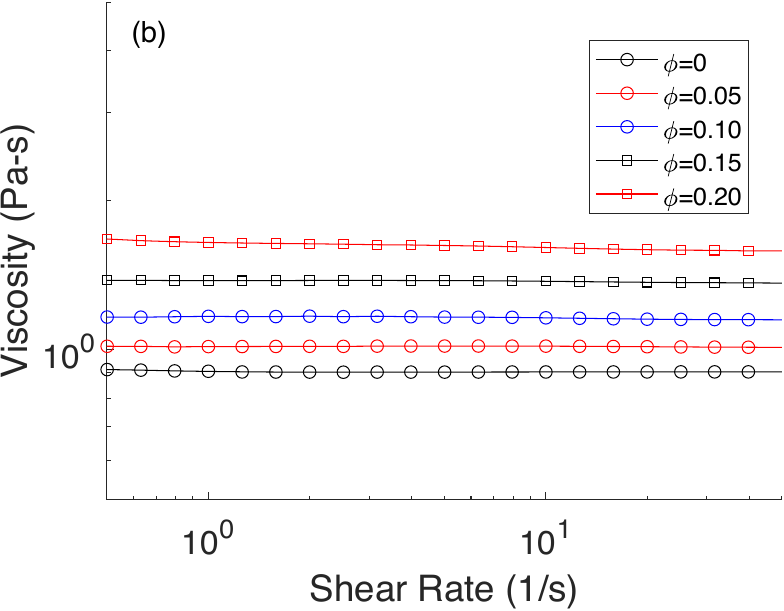}
    \includegraphics[width=0.32\textwidth]{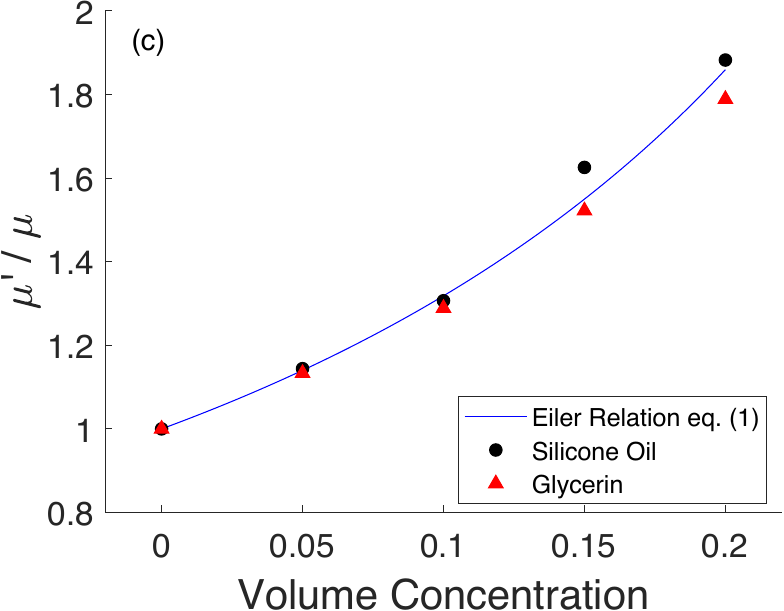}
    \caption{Characterization of fluids with different particle sizes. (a) Viscosity vs.~shear rate for Silicone oil DMS-T31 with $125 \, \si{\mu m}$ radius Polystyrene spheres. (b) Viscosity vs.~shear rate for Glycerin with $30 \, \si{\mu m}$ radius Poly(methyl methacrylate) spheres. Typical experimental shear rate is $\dot{\gamma} = \Omega R/L \approx 1 \, \mathrm{s}^{-1}$. (c) Comparison of effective viscosity vs.~volume concentration for experiments and Eiler relation in Eq.~\eqref{eq:eilerfit}~\cite{LinaresGuerrero2017}.}
    \label{fig:suspension_charac}
\end{figure}

To prepare the suspensions, we mix solid spherical particles with high-viscosity fluids and choose particle-fluid pairs such that the particles are neutrally buoyant. 
We measure the effective viscosities $\mu'$ of the $30$ and $125-\si{\mu m}$ suspensions and obtain that, for a given volume fraction, they do not depend on the imposed shear rate (see Fig.~\ref{fig:suspension_charac}). Note that the $300$-$\mu \textrm{m}$ suspension could not be characterized in the rheometer as the particles are approximately the same size as the concentric cylinder annular gap. 

For the $30$-$\mu \textrm{m}$ radius particles, poly(methyl methacrylate) (PMMA, $\rho_s = 1200 \; \mathrm{kg/m}^3$) spheres are suspended in liquid glycerin ($\rho = 1260 \; \mathrm{kg/m}^3, \; \mu = 0.90 \; \mathrm{Pa \; s}$). 
For larger particles $d = 125, \, 300 \, \si{\mu m}$, we use polystyrene ($\rho_s = 1000 \; \mathrm{kg/m}^3$) particles in liquid silicone oil DMS-T31 ($\rho = 970 \; \mathrm{kg/m}^3, \; \mu = 1.0 \; \mathrm{Pa \; s}$). 
The density $\rho_s$ of the solid particles is given by the manufacturer's Safety Data Sheet, and we assume a packing density of $64\%$ to obtain their volume. The density $\rho$ of the fluids is measured using a glass pycnometer at $22 \; ^\circ \mathrm{C}$. For each concentration, the suspension is mixed using 400 mL of fluid with a volume $V_\mathrm{solid}$ of particles so that $\phi = V_{\mathrm{solid}} \, / \, \left( V_\mathrm{solid} + V_\mathrm{fluid} \right)$ and characterized independently. 
The dynamic viscosities are obtained with a concentric cylinder rheometer (TA ARES G2). The characterization was done both in increasing and decreasing shear-rate order with similar results.

The effective viscosity of the suspension, $\mu '$, increases with the particle concentration $\phi$, as expected (Fig.~\ref{fig:suspension_charac}c); for a given particle fraction, it is independent of the imposed shear rate (Fig.~\ref{fig:suspension_charac}a-b). The viscosity dependence with $\phi$ is well reproduced by the empirical Eiler relation~\cite{kulkarni2008,zarraga2000}
\begin{equation} \label{eq:eilerfit}
\frac{\mu '}{\mu} = \left( 1 + \frac{a_e\phi}{1-\phi/\phi_m} \right)^2,
\end{equation}
with $a_e=1.25 \; \text{and} \; \phi_m = 0.64$, in agreement with previous experiments~\cite{LinaresGuerrero2017}.
Importantly, even though the particle radius is different in glycerin and silicone oil ($30 \, \si{\mu m}$ and $125 \, \si{\mu m}$, respectively,  the change in viscosity with particle concentration is essentially the same, as shown in Fig.~\ref{fig:suspension_charac}c. 
Our suspensions therefore allow us to explore the effect of the particle size and concentration on helical propulsion in suspensions with similar effective viscosities that are independent of shear rate.

\subsection{Force-torque measurements and propulsion efficiency} \label{subsec:results}
\subsubsection{Force and torque data}

\begin{figure}[t]
\centering
    \includegraphics[width=.95\textwidth]{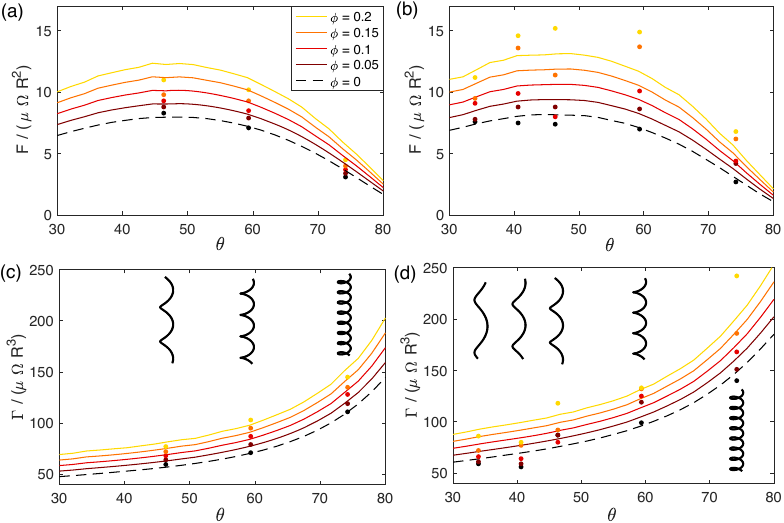}
    \caption{Normalized forces $F/(\mu \Omega R^2)$ (a,b) and torques $\Gamma/(\mu \Omega R^3)$ (c,d) as a function of the pitch angle $\theta$, with $\tan \theta = 2\pi R/ \lambda$, for several particle volume fractions $\phi$ (particles have a 30~$\mu$m radius). 
  Results in (a,c) show forces and torques on thin helices, $a/R=0.06$, and (b,d) on thicker ones with $a/R=0.13$. 
    Filled symbols represent experimental measurements, dashed lines represent the Newtonian slender-body theory, and solid lines represent the predictions from our modified slender-body theory simulations for suspensions. }
    \label{fig:force_torque_model}
\end{figure}

The experimental forces $F$ and torques $\Gamma$ at different particle concentrations $\phi$ are shown in Fig.~\ref{fig:force_torque_model} for various helical geometries: pitch length $\lambda$ and filament radius $a$. 
The forces and torques are normalized by $\mu \Omega R^2$ and $\mu \Omega R^3$ respectively, where the viscosity $\mu$ is taken to be that of the Newtonian fluid (not the effective viscosity). 
We find that increasing the dilute particle concentration systematically increases both the net force $F$ and torque $\Gamma$ along the axis of the helix for all geometries. This increase in the force and torque depends non-linearly on the geometry of the helix. 

\subsection{Estimating the propulsion efficiency} \label{subsec:RFT}

The net propulsion speed of a helical swimmer actuated at a constant angular velocity $\Omega$ depends on the relative increases of the torque and the force. 
To estimate the change in swimming speed from our experimental force and torque data, we now introduce the drag coefficient ratio $\xi$ inspired by classical results for resistive-force theory (RFT) in Newtonian fluids as an empirical proxy for helical propulsion efficiency.

The underlying idea of RFT is to approximate the forces on an elongated filament by a superposition of forces on small, locally-straight cylindrical elements~\cite{cox1970,lauga2020}. In the Newtonian limit, RFT has a rigorous mathematical basis in the asymptotic limit of slender rods~\cite{lauga2020}. 
For sufficiently slender ($a \ll \lambda$) and weakly bent ($\lambda \lesssim R $) filaments, the leading-order contribution to the hydrodynamic forces is proportional to the local flagellum velocity with drag coefficients (of proportionality) denoted by $\xi_{||}$ and $\xi_{\perp}$ along the tangential and normal directions, respectively.
Under these assumptions, the relationship between the local velocity $\mathbf u(s,t)$ of a segment of the filament at position $s$ and the hydrodynamic force (per unit length) $\mathbf{f} (s,t)$ acting on it is written
\begin{equation}
    \mathbf{f}(s,t) = - \left[ \xi_{||} \mathbf{t} \mathbf{t} + \xi_{\perp} \left( 1 - \mathbf{t} \mathbf{t} \right) \right] \cdot \mathbf{u}(s,t), 
\end{equation}
where $\mathbf t$ is the local unit tangent vector (allowed to vary spatially). 

Flagellar propulsion relies on drag anisotropy, meaning $\xi_{||} \neq \xi_{\perp}$, which results in possible misalignments between applied forces and resulting velocities. The force coefficient ratio is given by $\xi = \xi_{\perp} / \xi_{||}$; in the Newtonian case, $1 < \xi \leq 2 $, with $\xi \rightarrow 2$ the limit of an infinitely long straight filament.

For a rotating helix in a viscous fluid, drag anisotropy leads to a net propulsive force along its axis. Specifically, the propulsion speed $U$ of a force-free, torque-free helix of radius $R$ and pitch angle $\theta$ rotating at a constant rate $\Omega$ is given by~\cite{leshansky2009, lauga2020} 
\begin{equation}
    \frac{U}{\Omega R} = \frac{\left( \xi - 1 \right) \sin(2 \theta) }{\left( \xi + 1 \right) \left(1 + \sin ^2 \theta \right) } . 
    \label{eq:swim_predict}
\end{equation}
This relation shows that the drag coefficient ratio $\xi$ can be used as an empirical proxy for propulsion efficiency since the net velocity of the helical swimmer for a given angular velocity $\Omega$ is proportional to $(\xi -1)/(\xi + 1)$. 

The force and torque per unit length can also be expressed as a function of the tangential and normal drag coefficients~\cite{rodenborn2013}, with 
\begin{equation}
\begin{aligned}
        F &= \Omega R ( \xi_\perp - \xi_{||}) \sin \theta , \\
    \Gamma &= \Omega R^2 ( \xi_\perp \cos\theta + \xi_{||} \tan\theta \sin \theta) .
\end{aligned}
\label{eqFT}
\end{equation}
We therefore use our experimental force-torque data to obtain a force-torque drag coefficient ratio $\xi_{\textsc{F}\Gamma}$ that approximates the drag coefficient ratio $\xi$ by inversing equation~\ref{eqFT}
\begin{equation}
\xi_{\textsc{F}\Gamma} = \frac{\Gamma + FR \tan(\theta)}{\Gamma - FR \cot(\theta)}. 
\label{eq:xi_force_torque}
\end{equation}
In the idealized RFT framework, the two drag coefficient ratios are equal, and $\xi_{\textsc{F}\Gamma} = \xi$. Here, we extend the use of this coefficient $\xi_{\textsc{F}\Gamma}$ as a proxy for the efficiency of flagellar propulsion of our real helix, with an increase in $\xi_{\textsc{F}\Gamma}$ capturing an increase in distance traveled per rotation.

\begin{figure}[t]
    \centering
\includegraphics[width=.82\textwidth]{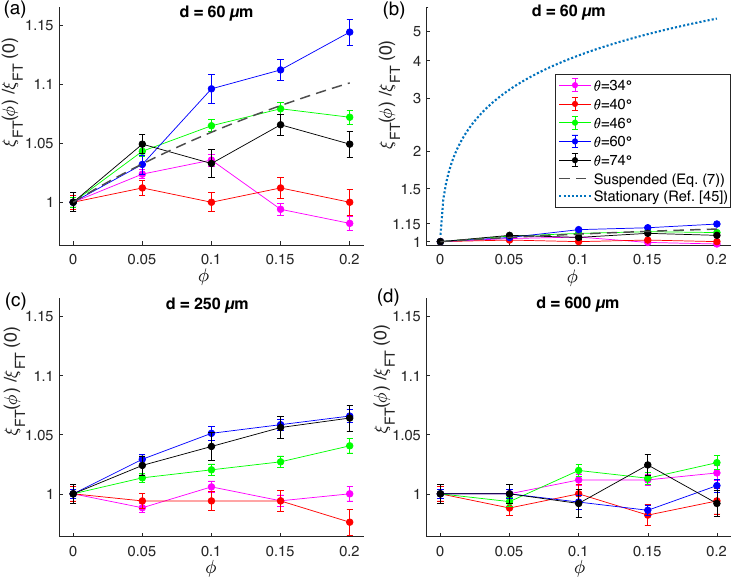}
    \caption{Change in coefficient ratio $\xi_{\textsc{F}\Gamma}(\phi)/\xi_{\textsc{F}\Gamma}(0)$ as a function of volume fraction $\phi$ for multiple particle radii $d$, (a,b) $d = 30\, \si{\mu m}$, (c) $d = 125\, \si{\mu m}$, (d) $d = 300\, \si{\mu m}$.; in all cases, $a/R=0.13$. Experiments: symbols. 
    The dashed and dotted lines show comparisons with theory: (a) prediction for suspended obstacles using resistive-force theory for an infinitely long cylinder (Eq.~\eqref{dphiRFT} with $\Lambda \rightarrow \infty$) (black dashed line) and (b) for stationary obstacles from Ref.~\cite{leshansky2009} (blue dotted line).}
    \label{fig:xi_vs_phi}
\end{figure}

How is the drag anisotropy affected by suspended particles in the surrounding fluid? 
We focus on the normalized ratio $\xi_{\textsc{F}\Gamma}(\phi)/\xi_{\textsc{F} \Gamma}(0)$ at different volume fractions $\phi$, with $\xi_{\textsc{F} \Gamma}(0)$ denoting the same ratio but in a Newtonian fluid (i.e.~with no particles). 
The experimental values obtained from Eq.~\eqref{eq:xi_force_torque} are plotted in Fig.~\ref{fig:xi_vs_phi} for the different particle sizes $d$ and helix geometries $\theta$ in the case $a/R = 0.13$; the data for thinner helices is provided in Appendix~\ref{sec:appendixthinhelix}.
We find that for small and intermediate ($d = 30, 125 \, \si{\mu m}$) particles and pitch angles $46^\circ$ and above, the drag ratio $\xi_{\textsc{F}\Gamma}(\phi)$ increases significantly with the dilute volume fraction $\phi$, by up to $15 \% $ at $\phi = 20 \%$ for $d = 30 \, \si{\mu m}$ and $\theta = 60^\circ$ (Fig.~\ref{fig:xi_vs_phi}a). 
The normalized drag coefficient ratio $\xi_{\textsc{F} \Gamma}/\xi_{{\textsc{F} \Gamma}(0)}$ varies non-monotonically with the pitch angle $\theta$, and helices with $\theta = 60 ^\circ$ are the most sensitive to the presence of the spheres at any volume fraction.

The drag anisotropy increase for a given volume fraction is less pronounced when the particles are larger relative to the filament thickness. 
When the particle radius $d$ approaches or exceeds that of the filament $a$, here when $d = 300 \, \si{\mu m}$, the increase in $\xi_{\textsc{F} \Gamma}/\xi_{{\textsc{F} \Gamma}(0)}$ essentially vanishes (Fig.~\ref{fig:xi_vs_phi}c) for all pitch angles tested.
Conversely, for thin helices with $a/R = 0.06$, the drag anisotropy increases slightly in the presence of particles, but the change is much less pronounced than for thicker ones, as expected from the force-torque data in Fig.~\ref{fig:force_torque_model}.

\section{Models for helical propulsion in suspensions} \label{sec:theory} 
\subsection{Resistive-force theory} \label{subsec:RFTtheory}

\subsubsection{Propulsion through a matrix of stationary obstacles}

Our experiments reveal a marked increase in the drag anisotropy, and hence in the efficiency of helical propulsion, in a suspension. 
Intuitively, we expect this change to originate in the hydrodynamic interactions between the helix and the suspended particles, as contacts are rare in dilute suspensions and prevented by lubrication interactions. 
Existing models for helical motion in heterogeneous media considered an array of obstacles embedded in a stationary matrix~\cite{leshansky2009, Chen2020}, where the flow is governed by the effective Brinkman medium approximation with permeability $\alpha^{-2} = 2 d^2/9 \phi$~\cite{Howells1974}. 
This model predicts a fivefold increase in the drag coefficient ratio for $\phi \geq 0.15 \%$, which is at odds with experimental results from our setup and with existing literature in suspensions~\cite{kamdar2022a}. 
The predicted increase with $\phi$ is highly non-linear even in dilute cases, and deviates strongly from our results, as shown in the dotted blue line~\cite{leshansky2009} of Fig.~\ref{fig:xi_vs_phi}b. 
The assumption in these past studies that the spheres are fixed and thus impose net forces (leading to stokeslet flows) on the fluid yields much stronger hydrodynamic interactions. Thus, it is necessary to build a theoretical approach that focuses on force-free suspended particles instead. 

In what follows, we present two models for helical propulsion in the presence of suspended particles: a fully analytical approach in the case of a cylinder and a computational approach applied to a helical geometry. In both cases, the flow perturbations from the small freely floating spheres are modeled as stresslets, which are signatures of force-free singularities. 

\begin{figure}[t]
    \centering
    \includegraphics[width = .95\columnwidth]{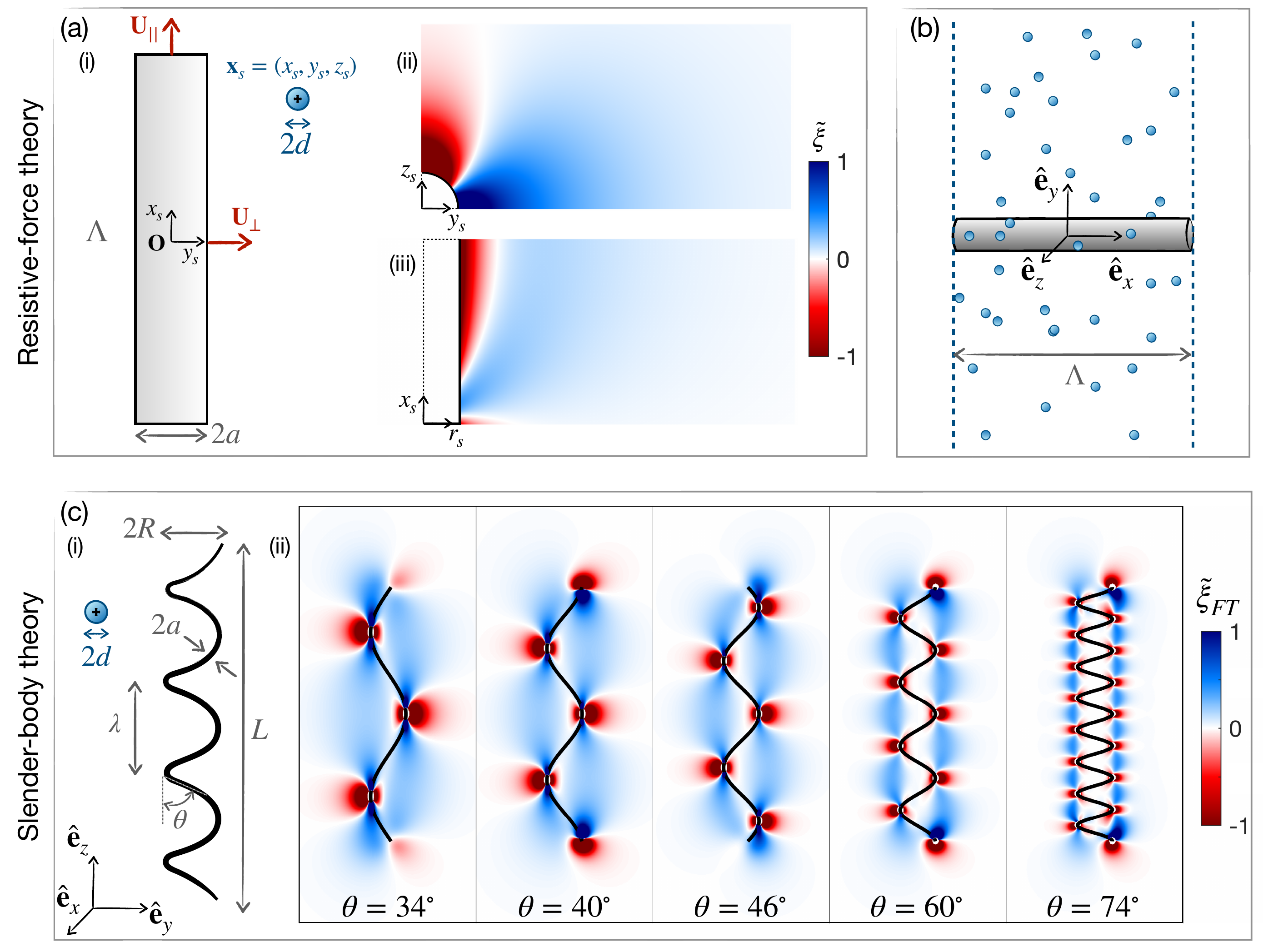}
    \caption{ 
    Modified resistive-force theory (top) and 
    slender-body theory (bottom) in the presence of suspended spheres.  
    (a) (ii) and (iii) Effect of the position of the sphere on the (rescaled) drag coefficient ratio of a cylinder $\tilde \xi $ along the (ii) radial (at the $x_s = 0$ plane) and (iii) longitudinal (averaged radially) directions. The colormap shows how the position of the sphere sets its contribution to the drag coefficient ratio computed analytically from resistive-force theory. Spheres located in blue regions result in an increase in the drag anisotropy and hence the propulsion efficiency, while red a decrease.
    (b) Modified resistive-force theory in a suspension.    
    (c) (i) Setup and notations for RFT in the presence of a single suspended sphere.
    (ii) Contribution $\tilde \xi_{\textsc{F}\Gamma} $ of a suspended sphere at a given location to the drag coefficient ratio of a helix, computed for different pitch angles using force-torque simulations and Eq.~\eqref{eq:xi_force_torque}. 
    Data are rescaled by a constant to fall in the $[-1;1]$ range, so the colorbar only shows relative effects. }  
    \label{contrib}
\end{figure}

\subsubsection{Interaction of a cylinder with a single freely suspended sphere}

First, we focus on the change in drag anisotropy for a cylinder and develop a modified resistive-force theory (RFT) that accounts for the hydrodynamic perturbation around a straight flagellum moving parallel and perpendicular to its axis in a suspension. 
In a Newtonian fluid, the flow induced by a moving cylinder can be approximated as a line of hydrodynamic singularities (stokeslets and source dipoles) allowing to compute the drag coefficients $\xi_{||}^{(0)}$ and $\xi_{\perp}^{(0)}$~\cite{lauga2020}. 
We now extend this approach to include, in the dilute limit and for small spheres, the disturbance flow caused by the presence of suspended spheres.

We first consider the effect of a single hard-sphere of radius $d$, located at $\mathbf{x}_s = (x_s, y_s, z_s)$ on the motion of cylinder ($-\Lambda/2 < x_c < \Lambda/2$) in the parallel and perpendicular directions (Fig.~\ref{contrib}a(i)). 
We compute the perturbation to the flow on the centerline of the cylinder in the presence of the small ($d \ll L$) free-floating sphere as the first reflection from the method of reflections~\cite{kim1991, guazzelli2011}. 
The unperturbed flow from the cylinder $\mathbf{u}_f$ is modeled as a line of stokeslets~\cite{lauga2020}, as the source dipoles decay faster and can be neglected here. 
The response induced by the free-floating sphere in this background flow is known to be a stresslet $S_{ij}$ of strength~\cite{guazzelli2011}
\begin{equation}
    S_{ij} (\mathbf{u}_f) =  \frac{20}{6} \pi \mu d^3 \left( \nabla \mathbf{u}_f + ^\intercal\!\nabla \mathbf{u}_f  \right). 
    \label{stresslet}
\end{equation}
Details of the calculations, and the analytical expressions of the modified drag parallel and perpendicular coefficients in the presence of a single sphere, $ \xi^{(0)}_{||} \left( 1 + \delta \xi_{||}(\mathbf{x}_s) \right) $ and $\xi^{(0)}_\perp \left( 1 + \delta \xi_{\perp}(\mathbf{x}_s) \right) $ respectively, are given in Appendix~\ref{sec:appendixRFT}.

We find that a single freely suspended sphere can either increase or decrease the drag coefficient ratio depending on its position relative to the cylinder. We map the contribution to the drag coefficient ratio of a sphere at a given location in Fig.~\ref{contrib}a. 
A sphere located in a blue region increases the drag anisotropy, while one in the red area on the side of the moving cylinder instead decreases it. 
Focusing on the effect in the $x_s = 0$ plane, we find in Fig.~\ref{contrib}a(ii) that a red region of decreased anisotropy develops on the side of the perpendicularly moving cylinder (along the $z_s$-direction) because a sphere located there increases the parallel drag coefficient, $\xi_{||}$ but, by symmetry, does not affect the perpendicular one $\xi_\perp$. On the other hand, spheres located in the front and back directions (blue region along the $y_s$-axis) experience a stronger flow from, and hence have a stronger response to, perpendicular motion, therefore increasing anisotropy (see Supplementary Material for details). 
Averaging the contributions radially in Fig.~\ref{contrib}a(iii) shows that the red regions (decreased anisotropy) dominate close to the cylinder, while the blue ones are stronger away from the helix.

\subsubsection{Cylinder in a dilute suspension}

In the case of a distribution of spheres, we next assume that the suspension is sufficiently dilute so that we neglect the screening hydrodynamic interaction between spheres. 
The modified drag for a suspension is then obtained by averaging the effect of a single sphere over all their possible positions, weighted by the volume fraction of spheres $\phi$. 
We take the spheres in the suspension to be uniformly distributed in space in the lateral region outside the cylinder, with $-\Lambda/2 < x_s < \Lambda/2 $, to avoid end effects at the tips of the cylinder (Fig.~\ref{contrib}b). We prevent overlap between the spheres and the cylinder by ensuring that the distance between the particles and the cylinder centreline is greater than the sum of their radii, namely $r_s > a+d$ where $r_s^2 = y_s^2+z_s^2$. 
The modified drag coefficients at volume fraction $\phi$ are then given by the superposition
\begin{equation}
    \xi_{\rm i }(\phi) = \xi_{\rm i }^{(0)} \left[ 1 +  \frac{\phi}{4/3 \pi d^3} \iiint_{r_s>(a+d), |x_s|<\Lambda/2} \delta \xi_{\rm i }(\mathbf{x}_s)  \dd \mathbf{x}_s \right],
\end{equation}
for $\rm{i} = ||, \, \perp$. 
The resulting drag coefficients can be computed analytically in the limit of a long cylinder, $ \lim_{\Lambda \to \infty} \xi_{\mathrm{i}}(\phi)$, by taking advantage of the symmetry in the $x$-direction. In that case, we obtain analytical expressions for the parallel and perpendicular drag coefficients as
\begin{equation}
\begin{split}
   \delta  \xi_{||}(\phi) & = \frac{ 5 \phi \xi_{||}^{(0)}  }{16 \pi \mu } \left[ 2 \log \left(\frac{\Lambda^2}{4(a+d)^2}+1\right)+\frac{-8 \Lambda^2 (a+d)^2-3 \Lambda^4}{\left(4(a+d)^2+\Lambda^2\right)^2} \right] , \\ 
    \delta \xi_{\perp}(\phi) & =  \frac{5 \phi \xi_{\perp}^{(0)} }{32 \pi \mu }  \left[   2 \log \left(\frac{\Lambda^2}{4(a+d)^2}+1\right)+\frac{16 \Lambda^2 (a+d)^2+3 \Lambda^4}{\left(4(a+d)^2+\Lambda^2\right)^2}\right] . 
    \label{dphiRFT}
\end{split}
\end{equation}

\subsubsection{Results: RFT predictions in a dilute suspension}

Our modified resistive-force theory in Eq.~\eqref{dphiRFT} predicts an increase in the drag anisotropy of a cylinder, and an improvement in helical propulsion, in a suspension. 
We compare these predictions from Eq.~\eqref{dphiRFT} with the experimental data in Fig.~\ref{fig:xi_vs_phi}a, taking the bulk drag coefficients for the surrounding fluid alone to be $\xi_{||} = 2 \pi \mu / \log(2L/a)$ and $\xi_\perp = 2 \xi_{||}$. 
Quantitatively, these predictions are consistent with the experimental drag anisotropy enhancement, thus providing an intuitive physical explanation for our experimental observations. 
However, the quantitative results are quite sensitive to the dimensions of the cylinder, radius $a$ and length $\Lambda$, and, critically, to the choice of values for the Newtonian drag coefficients $\xi_{||}^{(0)} $ and $\xi_{\perp}^{(0)} $ for a finite cylinder~\cite{rodenborn2013}, for which several expressions have been proposed in the literature~\cite{gray1955,lighthill1976}.

\subsection{Slender-body theory} \label{subsec:SBT}

Non-local hydrodynamic interactions between different segments of the helix lead to significant corrections to RFT, which we now consider~\cite{lauga2020}. Hydrodynamic interactions with the spheres will then depend both on their position relative to the helix, but also on the geometry of the helix itself. 
In this section, we extend the above RFT approach to Lighthill's slender-body theory (SBT)~\cite{lighthill1976}, allowing us to make quantitative predictions on our experimental results. 

\subsubsection{Model and implementation}

In the bulk fluid, Lighthill's SBT relates the local velocity of a segment of the helix at arclength $s$, $\mathbf{u}_h(s)$, and the local force per unit length $\mathbf{f}_h(s)$ in the form 
\begin{equation}
\label{SBTL}
    \mathbf{u}_h(s) = \frac{1}{4 \pi \mu} (\mathbf{I} - \mathbf{t}\mathbf{t}) \cdot \mathbf{f}_h (s) + \frac{1}{8 \pi \mu} \int_{r \geq \delta } \left(\frac{\mathbf{1}}{r} + \frac{\mathbf{r}\mathbf{r}}{r^3} \right) \cdot \mathbf{f}_h (s') \dd s',
\end{equation}
where $\delta = a \sqrt{e} /2$. 
When the rotation $\Omega$ of the helix is imposed along its axis and no translation occurs, $\mathbf{u}_h(s)$ is known everywhere. We implement this numerically by discretizing the helix in edges of length $2 \delta$~\cite{jawed2017}. The flow from the helix comes from stokeslets located at each node of the discretized helix, and we obtain their strength $\mathbf{f}_h$ numerically. Our simulations are validated against the classical Newtonian SBT~\cite{lighthill1976,rodenborn2013}. Details on the discretization method are included in Appendix~\ref{sec:appendixSBT}.

If a suspended sphere is located at $\mathbf{x}_s$, it experiences a flow $\mathbf{u}_h(\mathbf{x}_s)$ which is the sum of the above stokeslet from each node of the moving discretized helix. 
In turn, the sphere generates a reflection flow given by the stresslet $\mathbf{S}$ from Eq.~\eqref{stresslet}. The disturbance flow at a node on the helix $\mathbf{x}_0$ in the presence of a single sphere writes 
\begin{equation}
    {u}_{s,i}^p (\mathbf{x}_0) = - \frac{ 3}{8 \pi \mu }  S_{jk} ( \mathbf{u}_h (\mathbf{x}_s) )  \frac{  \tilde{x}_{0i}^p \tilde{x}_{0j}^p \tilde{x}_{0k}^p } { | \tilde{\mathbf{x}}_0 ^p |^5}  ,
\end{equation}
with $\tilde{\mathbf{x}}_0 = \mathbf{x}_0 - \mathbf{x}_s $. 

For a suspension, we draw spheres randomly around the helix. 
We avoid overlap between the sphere and the helix, 
we set a minimum distance between the helix centerline and the particle center $d_{min} = a+d$.
We also ensure that the integration is independent of the size of the integration domain that contains the spheres and the number of spheres. 
Finally, assuming that the suspension is dilute, we average the contributions of individual particles and scale the resulting forces and torques with the volume fraction $\phi$. 

\subsubsection{Numerical results}

With this computational model, we can reproduce the experimental set-up where the helix rotates at imposed rate $\Omega$ without moving ($U = 0$) and perform a new force balance using the modified local velocities on the helix to obtain the total force $F$ and torque $\Gamma$ along the axis of the helix. We compare these predictions (solid lines) to the experimental data (filled circles) in Fig.~\ref{fig:force_torque_model}. 

In all cases, we find that the presence of suspended spheres in the far field consistently enhances helical propulsion efficiency for the range of helical geometries $\lambda$ that we explore. 
For small spheres ($d = 30\, \si{\mu m}$), we find a good quantitative agreement between the force and torque from our experimental data and our modified SBT. 
However, for thinner helices ($a/R = 0.06$), our predictions slightly overestimate the increased propulsion force $F$, while they underestimate it in some geometries for thicker helices ($a/R = 0.133$). 
The discrepancy at larger thicknesses could be due to the decreased slenderness $a/R$ of the helical radius, a central hypothesis for SBT. This leads the model to underestimate the change in $\xi_{F \Gamma}$. 

Overall, our slender-body theory model confirms our simplified results focusing on a cylinder: far-field hydrodynamic interactions with freely suspended spheres can significantly increase swimmer propulsion. In addition, the slender-body theory is more precise as it allows us to understand how the enhancement affects the force and torques on the helix. It also reproduces the experimental results' dependence on the geometry of the swimmer.


\subsubsection{Influence of the position of the spheres}

To get further physical insight, and as in the case of RFT for a cylinder, we now investigate the effect of the spatial location of the spheres relative to the helix in SBT. We perform slender-body simulations in the presence of a single sphere and compute the corresponding drag coefficient ratio. 
A single sphere has a very weak effect, but, as for a cylinder, it can either increase or decrease the drag coefficient ratio depending on its position. 

In Fig.~\ref{contrib}c, we map how a sphere at a given position contributes to the helix drag coefficient ratio. 
The colormap shows the contribution $\tilde \xi_{\textsc{F}\Gamma}$ for spheres located on a plane containing the axis of the helix.
Spheres in a helical region surrounding the helix decrease the drag coefficient ratio (red regions in Fig.~\ref{contrib}c), while others increase it (blue regions). 
This mirrors the case of a cylinder, in which spheres on the $y_s = 0$ sides of the moving cylinder decrease the drag anisotropy, with additional complexity in the contributions from spheres at different positions stemming from the geometry of the helix. 
Our model shows that the propulsion efficiency of the helix is sensitive to the spatial distribution of the spheres. Because positive and negative contributions balance, long-range effects from spheres at several helical radii $R$ away also become important, and the system depends critically on the distribution of suspended particles. 

\section{Propulsion of a free swimmer}
\label{sec:swimmers}

Equipped with our understanding of the force generation in sections~\ref{sec:experiment} (experiments) and~\ref{sec:theory} (modeling), we now test our predictions against the locomotion of a force-free artificial swimmer propelled by a rigid helix. 

\subsection{Propulsion of a force-free artificial microswimmer in a suspension}

\subsubsection{Experimental setup}
\begin{figure}[t]
    \centering
    \includegraphics[width = \columnwidth]{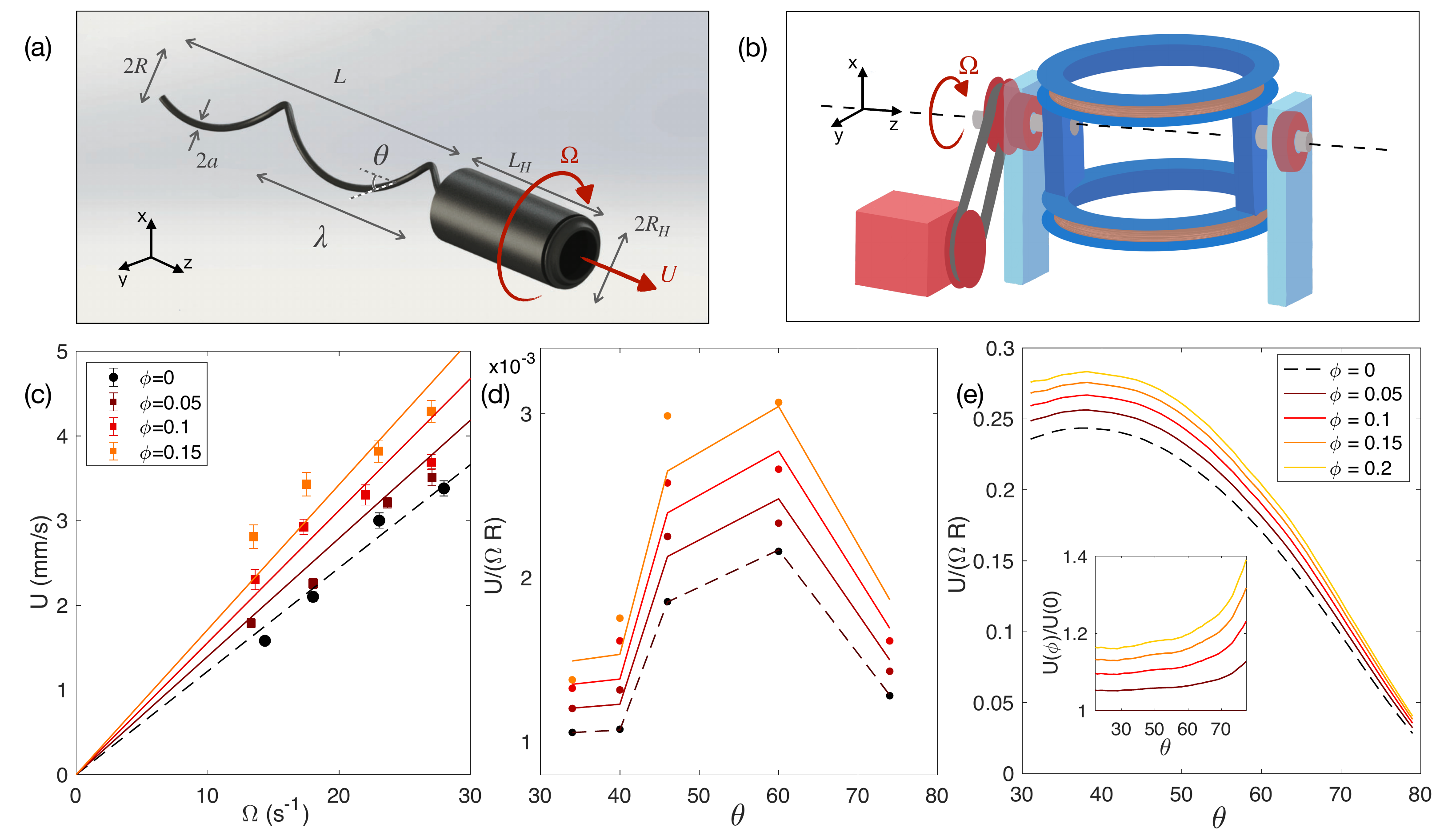}
    \caption{Experiments and predictions for the speed of a force-free helical swimmer moving through a suspension. (a) Design of the swimmer, with a helical tail and a cylindrical head containing a magnet. (b) The tank containing the swimmer is placed between rotating Helmholtz coils actuated by a motor.   
    (c) Raw data for the force-free swimming speed $U$ in experiments with $\theta = 60^\circ$ for different particle volume fractions $\phi$ and angular velocities $\Omega$. The dashed line at $\phi = 0$ shows the results from classical bulk SBT simulations with an additional fitted drag force from the head. The solid lines show the speed increase predictions in the presence of suspended spheres from our modified SBT simulations. 
    (d) Experimental rescaled force-free swimming velocities, with our modified suspension SBT predictions for the speed increase in suspensions. The dashed black line shows the fit to the Newtonian case. 
    (e) Predicted velocity of a free-swimming helix with no head using slender-body theory in the bulk fluid (dashed line, $\phi = 0$) and in suspensions (solid lines, $\phi = 0.05 - 0.2$). The inset shows the relative increase in a suspension $U(\phi)/U(0)$ depending on the geometry.  } 
    \label{fig:freeswimmer}
\end{figure}

Our experimental setup consists of a force-free swimmer with a cylindrical head and a helical tail that is externally actuated by a rotating magnetic field. 
As a result, a constant angular velocity $\Omega$ is imposed along its axis, in the $z$-direction, and we measure the translation velocity of the swimmer along its axis $U(\phi)$ for different concentrations of suspended particles (Fig.~\ref{fig:freeswimmer}a). 
The swimmer moves through a fluid-filled tank placed inside a pair of rotating Helmholtz coils, as shown in Fig.~\ref{fig:freeswimmer}b, building on an existing setup~\cite{godinez2012}. We obtain the translational velocity resulting from its force-free motion along the $z$-direction from videos. 

The swimmers are 3D-printed (Formlabs Form 3B) as one body without a cap on the cylindrical head, and a magnet is placed inside the body, aligned with the magnetic field from the Helmholtz coils. The head then is filled with the surrounding fluid such that the swimmer is neutrally buoyant, and sealed with a threaded cap. The swimmer is shown schematically in Fig.~\ref{fig:freeswimmer}a. The helix has a length $L= 40$ mm, radius $R=2.25$ mm and filament radius $a=0.3$ mm ($a/R= 0.13$). The head dimensions are $L_H=13$ mm and $R_H=2.25$ mm. 

The fluids used in these experiments are the suspensions with 30 $\si{\mu m}$-radius particles described in section~\ref{sec:experiment}.
The rotation of the coils is controlled by applying a consistent voltage to a motor connected to the coils. The voltage-velocity relation is characterized by using a tachometer. 
Since the head of the swimmer contains a magnet, the entire swimmer rotates with the magnetic field at the prescribed angular velocity $\Omega$. 
We ensure that the propulsion occurs at a low Reynolds number by setting a maximum $\Omega$ so that $Re = \rho \Omega R^2 / \mu < 0.1$, with $\rho$ and $\mu$ the density and viscosity of the bulk fluid. 
To record time and capture images that are not obstructed by the rotation of the coil, a photo-detection system is added that sends pulse signals to our camera (OpenMV CAM H7) only when the coils are not blocking the camera. 

\subsubsection{Speed increase for a free swimmer: experiments}
The measured swimmer speeds in a suspension are shown in Fig.~\ref{fig:freeswimmer}c,d. 
First, we measure the swimming velocity $U$ for a given pitch angle at different angular velocities $\Omega$, with results for $ \theta = 60 ^\circ$ plotted in Fig.~\ref{fig:freeswimmer}c; swimming speeds for the remaining pitch angles ($\theta = \{34{^\circ}; 40{^\circ}; 46{^\circ}; 74{^\circ}\}$) are shown in Appendix~\ref{sec:appendixswimmingall}. Because the velocity scales linearly with $\Omega$ at low Reynolds numbers, we then extract the rescaled velocity $U/(\Omega R)$ using a least-square method with linear relationship, which we plot for all the tested geometries in Fig.~\ref{fig:freeswimmer}d. 
Importantly, all swimmers are seen to move faster in dilute suspensions than in Newtonian fluids. 
The swimmers with pitch angles \ang{46} and \ang{60} are the fastest regardless of $\phi$, in agreement with the angles displayed by flagellated microorganisms as well as with theoretical predictions~\cite{lauga2020}.
The speed increase in suspensions also depends strongly on geometry, with the largest increases for $\theta = 40, 46^\circ$, of over $60\%$ when $\phi = 0.15$ against $ \approx 30\%$ for $\theta = 34^\circ$. 

\subsection{Slender-body theory for a free swimmer}

We now extend the slender-body theory (SBT) model developed above to the case of a free-swimming helix actuated with a constant angular velocity $\Omega$. In the case without a head, we impose that the swimmer remains force-free along its axis (i.e.~$F = 0$ in the $z$-direction) to predict the resulting swimming speed $U$. 
We plot the swimming speed $U$ in Fig.~\ref{fig:freeswimmer}e. All helices translate faster, with increases relative to the bulk fluid speed $U(0)$ ranging from $\approx 10 \%$ to $30\%$ when $\phi = 0.2$. The strongest relative increase ($> 30\%$) occurs for $\theta = 70^\circ$, but the fastest helix is still the optimal one around $45 ^\circ$. 

Finally, we extend our model to the case of a swimmer with a head by including an additional drag in the $z$-direction to account for the drag on the head. 
As a minimal model, we chose to include only a constant drag coefficient, independent of the fluid, which we fit for each geometry to our experimental data in the Newtonian case, as shown by the dashed black line in Fig.~\ref{fig:freeswimmer}d. 
We then obtain speed increase predictions by performing our SBT simulations in suspensions and adding this additional fitted drag to the force balance. The predictions are shown in solid colored lines in Fig.~\ref{fig:freeswimmer}d. As expected, the additional drag means that the speed of a swimmer with a head is lower than that of an isolated helix, but it still increases significantly in a suspension (as compared to locomotion in a bulk fluid), in quantitative agreement with experimental results.

\section{Discussion} \label{sec:discussion}
\subsection{Summary of our work}
Our combined experimental-theoretical study reveals that enhancement of helical propulsion and locomotion in a suspension is a very robust physical feature, both for an isolated helix and for a swimmer with a head, and that it depends on the concentration of particles, helix pitch angle, particle size, and helix radius relative to filament radius. 

Our first experimental setup measures the forces and torques on a rotating helix in a suspension and uses them to estimate its propulsion efficiency. 
We then study the propulsion theoretically in suspensions by extending two classical modeling approaches, resistive-force theory (RFT) and slender-body theory (SBT), to the case of dilute suspensions of freely suspended spheres. We quantify analytically and numerically the hydrodynamic interaction between a collection of suspended spheres and a moving cylinder and helix, modeled as lines of hydrodynamic singularities. 
The analytical RFT qualitatively captures the propulsion enhancement at a set angular velocity. The numerical SBT allows us to make quantitative predictions that agree with our experiments for a range of helix geometries and particle concentrations, provided that the filament radius remains thin relative to the other helical length scales and the particles are small. 
Our work also demonstrates the limitations of using fixed resistance models for porous media~\cite{leshansky2009} to account for obstacles that are suspended in a fluid.
Strikingly, we also show that depending on the position of the suspended spheres, they can contribute positively or negatively to the drag coefficient ratio. This highly non-linear effect of the spheres underlines the complexity of the system's dynamics, which cannot be straightforwardly reduced to a continuous model. 

We next test our experimental and theoretical predictions by studying force-free artificial swimmers actuated by an external rotating field. All the swimmers studied move faster through the suspension compared to a Newtonian fluid given a prescribed angular velocity, with speed increases of up to $60\%$ even in dilute suspensions. 
These experimental results are in qualitative agreement with our theoretical predictions for a free swimming helix. A similar agreement is obtained for the swimming speed of the swimmers by accounting for the drag force on the head of the swimmer. 

Our work shows that the presence of suspended obstacles should be accounted for when studying biological and artificial swimmers in complex environments. We provide a framework to study and predict the changes in propulsion efficiency at a given angular velocity. 
 This work could be extended to wave-based propagation or flexible flagella, as well as more complex environments combining complex rheologies and suspended obstacles~\cite{Spagnolie2023}, making it relevant to many biological and medical applications.

\subsection{Comparison with past studies}

\begin{figure}[t]
    \centering
    \includegraphics[width = 0.6\columnwidth]{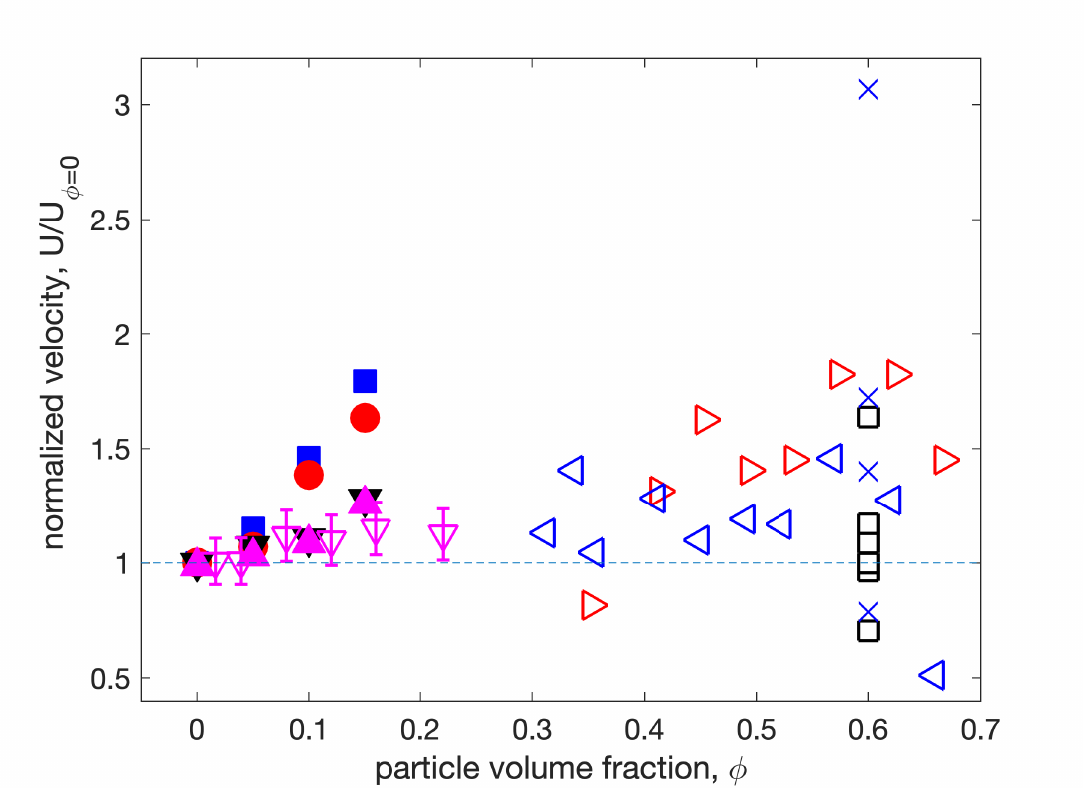}
    \caption{Normalized swimming speed, $U/U_{\phi=0}$, as a function of particle volume fraction, $\phi$. Data from this study: ({\color{blue} $\blacksquare$}, {\color{red} \LARGE{$\bullet$}}, {\color{magenta} \large{$\blacktriangle $}}, {\large{$\blacktriangledown$}}) for  $\theta=60^o$ and different angular velocities $\Omega$. Data from~\cite{Juarez2010}: ({\color{red} \Large{$\triangleright$}},{\color{blue} \Large{$\triangleleft$}}) for \textit{C.~elegans} in a monolayer of mono-dispersed glass particles . Data from~\cite{jung2010}: ({\color{black} $\square$}) for \textit{C.~elegans} in a quasi-two-dimensional granular slurry, for different particle sizes and $\phi\approx 0.6$ . Data from~\cite{park2016}: ({\color{magenta} \large{$\triangledown$}})
    \textit{C.~elegans}  in a shear-thinning colloidal suspension. Data from~\cite{kudrolli2019}: ({\color{blue} $\times$}) for \textit{Lumbriculus variegatus} (aquatic worm) in a packed clear hydrogel particle bed.
    }
    \label{fig:comparison}
\end{figure}

Given this new understanding, we are now in a position to compare our swimming results with other similar measurements from the recent literature. We show in Fig.~\ref{fig:comparison} the free swimming speed, $U$, normalized by its value in the absence of particles (i.e.~in the bulk fluid with $\phi$=0) for swimmers with different helix pitch angles (see Fig.~\ref{fig:freeswimmer}). As discussed above, the swimming speed always increases with particle concentration. For the range of concentrations that we tested (up to $\phi=0.15$), the swimming speed increased up to 1.8 times the speed without particles, for the maximum angular velocity ($\Omega \approx 28$ s$^{-1}$).

To our knowledge, no other experiments for helical force-free swimmers have been reported. Several studies have investigated the swimming speed of \textit{C.~elegans} nematodes in particulate media~\cite{Juarez2010,jung2010,park2016}. More recent work focused on experimentally characterizing  
the burrowing dynamics of aquatic worms in particle-saturated media~\cite{kudrolli2019}. We plot in Fig.~\ref{fig:comparison} the speed changes from these papers as a function of the particle volume fraction $\phi$ (empty symbols), and compare this data to our results for helical propulsion (filled symbols).  

Although the swimming strategy for these organisms is clearly different from the rotating helix considered in our study, many similarities are apparent. In most cases, the presence of particles leads to an increase in speed with respect to the swimming speed without particles. Also, the swimming speed appears to increase with particle loading, at least for the case of \textit{C.~elegans}. Note that in most of these other studies, the particle concentration is higher than $\phi=0.3$. For such concentrated suspensions, the particles are likely to make physical contact with the swimmer and hydrodynamic interactions may not be as important as in the present case. Such direct particle contact, which is not considered in the present study, may lead to an increase in the swimming speed of a different origin. Note that in one specific study~\cite{park2016}, swimming was studied for particle concentrations comparable to those considered here. In that case, an increase of the normalized swimming speed was reported but for colloidal suspensions, which are known to exhibit shear-thinning rheology which, in turn, may explain the increase in swimming speed~\cite{Gomez2016}.

\pagebreak

\begin{acknowledgments}
 A.T. received support from the Simons Foundation through the Math + X grant awarded to the University of Pennsylvania.
\end{acknowledgments}


\bibliography{bibhelix}

\newpage

 \setcounter{section}{0}
\appendix


\section{Modified resistive-force theory in a dilute suspension} \label{sec:appendixRFT}

We detail in this section the modified resistive-force theory (RFT) used to compute the drag coefficient ratio for a cylinder in a suspension $\xi (\phi)$. 

\subsection{Interaction between a cylinder and a single sphere}
\subsubsection{Suspended sphere in the background flow of a stokeslet}

We compute the disturbance flow from a suspended sphere of radius $d$ at $\mathbf{x}_s$ when a singularity stokeslet is placed at $\mathbf{x}=\mathbf{0}$. 
The sphere creates a stresslet located at $\mathbf{x}_s$ of strength $S_{ij} =  \frac{20}{3} \pi \mu d^3 \left\{ 1 + \frac{d^2}{10} \nabla ^2 \right\} e_{ij}$~\cite{guazzelli2011} with $\mathbf{e}$ is the rate of strain tensor of the ambient flow. Here, we consider small spheres ($d \ll L $) and neglect the Laplacian term. 
The velocity and rate of strain for a stokeslet at the origin are 
\begin{equation}
     \mathbf{u_s} = \frac{\mathbf{s}}{8 \pi \mu r} + \frac{(\mathbf{s}\cdot\mathbf{x}) \mathbf{x}}{ 8 \pi \mu r^3}
\quad \textrm{and} \quad 
    e_{ij} =  \frac{(\mathbf{s}\cdot\mathbf{x}) }{8 \pi \mu r^3} \delta_{ij} - \frac{ 3(\mathbf{s}\cdot\mathbf{x}) }{8 \pi \mu r^5} x_i x_j . 
\end{equation}
The presence of a suspended sphere creates a stresslet 
\begin{equation}
    S_{ij}(\mathbf{x}_s, \mathbf{s})   =  \frac{20}{3} \pi \mu d^3  e_{ij}(\mathbf{x_s})  = \frac{5}{6} d^3   \left\{ \frac{(\mathbf{s}\cdot\mathbf{x_s}) }{r_s^3} \delta_{ij} - \frac{ 3(\mathbf{s}\cdot\mathbf{x_s}) }{r_s^5} x_{s,i} x_{s,j} \right\} 
    \label{stokesletS}. 
\end{equation}
This produces the disturbance flow
\begin{equation}
\label{eqspeedsph}
\begin{split}
     u_{s,i}(\mathbf{x}_s,\mathbf{x}) & = \frac{-3 S_{jk}}{8 \pi \mu}  \frac{\tilde{x}_j \tilde{x}_k \tilde{x}_i}{\tilde{r}^5 } \\
     & = - \frac{5}{16 \pi \mu} d^3 (\mathbf{s}\cdot\mathbf{x_s})  \left\{ \frac{ (x_j - x_{s,j})^2  }{r_s^3}  - \frac{ 3 x_{s,j} x_{s,k} (x_j - x_{s,j}) (x_k - x_{s,k}) }{r_s^5}  \right\} \frac{(x_i - x_{s,i})  }{ \abs{\mathbf{x}-\mathbf{x}_s}^5 } . 
\end{split} 
\end{equation}

\subsubsection{Suspended sphere in the background flow of a cylinder moving along its long-axis}

The flow from a cylinder of length $\Lambda$ moving parallel to its axis at a set speed $\bm{U} = U \bm{\hat{e}}_x $ is approximated by a line of point forces with force density $\bm{F}_{||} /L = \xi_{||}^{(0)} U  \mathbf{e}_x$ for $-\Lambda/2 < x < \Lambda/2$ (see Fig.~\ref{contrib}a). 
In the absence of spheres, and in the limit of an asymptotically slender filament $a/\Lambda \rightarrow \infty$~\cite{lauga2020}, the drag coefficients are given by 
\begin{equation}
    \xi_{||}^{(0)} = \frac{2 \pi \mu}{\log \left(\Lambda /a \right)} \quad \textrm{and} \quad \xi_{\perp}^{(0)} = \frac{4 \pi \mu}{\log \left(\Lambda /a \right)}. 
\end{equation}

The flow at a point $s$ on the parallel cylinder induced by the presence of a freely suspended sphere at $\mathbf{x}_s$ is
\begin{equation}
    \mathbf{u}^{(s)} (s)  = \int_{-\Lambda/2}^{\Lambda/2} \mathbf{u}_s(\mathbf{x_s} - t \mathbf{e}_x, (s-t) \mathbf{e}_x) \dd t ,
\end{equation}
which we decompose using Eq.~\eqref{stokesletS} as
\begin{equation}
\begin{split}
    u_{||,x}^{(s)} (s) = & -  \frac{ 5 \xi_{||}^{(0)} U d^3}{16 \pi \mu } \int_{-\Lambda/2}^{\Lambda/2}  (x_s - t) \frac{(s-x_s)}{[(x_s-s)^2 +r_s^2]^{5/2}} \\
    & \left[ \frac{(s-x_s)^2 + r_s^2}{[(x_s-t)^2 +r_s^2]^{3/2}} - \frac{3 \left( (x_s-t)^2(s-x_s)^2 +2 (x_s-t)r_s(s-x_s)(-r_s) + r_s^2 (-r_s)^2 \right) }{[(x_s-t)^2 +r_s^2]^{5/2}}   \right]  \dd t \\
    = &  - \frac{ 5 \xi_{||}^{(0)} U d^3}{16 \pi \mu }
     \left\{ \frac{s-x_s} {[(x_s-s)^2+r_s^2]^{3/2}} \int_{-\Lambda/2}^{\Lambda/2} \frac{ x_s - t} {[(x_s-t)^2 +r_s^2]^{3/2}} \dd t   \right. \\
     & \qquad \qquad \quad + \frac{s-x_s} {[(x_s-s)^2+r_s^2]^{5/2}} \int_{-\Lambda/2}^{\Lambda/2} \frac{- 3 r_s^4  (x_s - t)} {[(x_s-t)^2 +r_s^2]^{5/2}}  \dd t  \\
     & \qquad \qquad \quad + \frac{(s-x_s)^2} {[(x_s-s)^2+r_s^2]^{5/2}} \int_{-\Lambda/2}^{\Lambda/2}  \frac{ 6 (x_s-t)^2 r_s^2} {[(x_s-t)^2 +r_s^2]^{5/2}}  \dd t\\
    & \qquad \qquad \quad + \frac{(s-x_s)^3} {[(x_s-s)^2+r_s^2]^{5/2}} \int_{-\Lambda/2}^{\Lambda/2}  \left. \frac{-  3 (x_s-t)^3} {[(x_s-t)^2 +r_s^2]^{5/2}} \dd t \right\} . 
    \label{eqsuspar}
\end{split}
\end{equation}

We rescale the position of the sphere by the cylinder length $\Lambda$,  so that $x_s \rightarrow x_s/\Lambda$, $r_s \rightarrow r_s/\Lambda$, 
and write
\begin{equation}
    J_{n,p}(x_s) =  \int_{-1/2}^{1/2} \frac{(x_s - t)^n} {[(x_s-t)^2 +r_s^2]^{p/2}} \dd t .
\end{equation}
We now assume that the modification of the cylinder speed due to the presence of the sphere is the average of the local modification over its length, and take advantage of the symmetry between $s$ and $t$ to obtain 
\begin{equation}
\begin{split}
    u_{||,x}^{(s)} (\mathbf{x}_s) &=  \frac{ - 5 d^3 \xi_{||}^{(0)} U}{ 16\pi \mu \Lambda^3} \left[- J_{1,3}^2 +3 r_s^4 J_{1,5} + 6 r_s^2 J_{2,5}^2 +3 J_{3,5}^2 \right]\\
    &=  \frac{ -5 d^3 \xi_{||}^{(0)} U}{ 16\pi \mu \Lambda^3} \frac{4}{3 r_s^2} \left[ 1 + \frac{1 -12 (r_s^2+x_s^2) + 48 (x_s^4-r_s^4+2r_s^2x_s^2) -64(r_s^2+x_s^2)^3}{\left[1+8 \left(r_s^2-x_s^2\right)+16 \left(r_s^2+x_s^2\right)^2\right]^{3/2} }\right] .
\end{split}
\end{equation}

The force balance on the cylinder writes $\xi_{||} ^{(0)} (\mathbf{U} + \bm{u}_{||}^{(s)} (\mathbf{x}_s)) = \xi_{||}(\bm{x}_s) \mathbf{U}$, meaning that the driving force compensates for the above flow from the sphere. The projection on the x-axis yields the modified drag coefficient, 
\begin{equation}
\begin{split}
    \xi_{||}(r_s,{x}_s) & = \xi_{||}^{(0)} \left[ 1 + \frac{5 \xi_{||}^{(0)} d^3}{12\pi \mu \Lambda^3}  \frac{1}{r_s^2}  \left( 1 + \frac{1 -12  (r_s^2+x_s^2) + 48  (x_s^4-r_s^4+2r_s^2x_s^2) -64(r_s^2+x_s^2)^3}{\left[1+8  \left(r_s^2-x_s^2\right)+16 \left(r_s^2+x_s^2\right)^2\right]^{3/2} }\right)  \right] \\
    & =  \xi_{||}^{(0)} \left( 1 + \delta \xi_{||}(\mathbf{x}_s) \right) . 
\end{split}
\end{equation}

\subsubsection{Perpendicular motion of cylinder}

We carry out the same analysis for a cylinder of length $\Lambda$ moving at a set speed $\bm{U} = U \bm{\hat{e}}_y $ perpendicular to its length. The flow from the cylinder can be described as a line of stokeslets with force density $ \xi_{\perp}^{(0)} U  \mathbf{e}_y$ and source dipoles of strength $ \frac{a^2}{4 \mu} \xi_{\perp}^{(0)} U  \mathbf{e}_y$  for $-\Lambda/2 < x < \Lambda/2$~\cite{lauga2020}. As the source dipoles decay faster, we find that their contribution to the drag coefficient change can be neglected. 

We again look for the local velocity at a point $s$ on the perpendicular cylinder from asphere at $\mathbf{x}_s$
\begin{equation}
\begin{split}
    \mathbf{u}^{(s)}_\perp (s) & = \int_{-\Lambda/2}^{\Lambda/2} \mathbf{u}_s(\mathbf{x_s} - t \mathbf{e}_x, (s-t) \mathbf{e}_x) \dd t .
\end{split}
\end{equation}
The component along the swimming direction writes
\begin{equation}
\begin{split}
     u^{(s)}_{\perp,y} (s) = & -  \frac{ 5 \xi_{\perp}^{(0)} U d^3}{16 \pi \mu } \int_{-\Lambda/2}^{\Lambda/2}  \frac{ - y_s^2 }{[(x_s-s)^2 +r_s^2]^{5/2}} \\
    & \left[ \frac{(s-x_s)^2 + r_s^2}{[(x_s-t)^2 +r_s^2]^{3/2}} - \frac{3 \left( (x_s-t)^2(s-x_s)^2 +2 (x_s-t)r_s(s-x_s)(-r_s) + r_s^2 (-r_s)^2 \right) }{[(x_s-t)^2 +r_s^2]^{5/2}}   \right]  \dd t \\
    = &  - \frac{ 5 \xi_{||}^{(0)} U d^3}{16 \pi \mu }
     \left\{ \frac{-y_s^2} {[(x_s-s)^2+r_s^2]^{3/2}} \int_{-\Lambda/2}^{\Lambda/2} \frac{ 1 } {[(x_s-t)^2 +r_s^2]^{3/2}} \dd t   \right. \\
     & \qquad \qquad \quad + \frac{-y_s^2 } {[(x_s-s)^2+r_s^2]^{5/2}} \int_{-\Lambda/2}^{\Lambda/2} \frac{- 3 r_s^4 } {[(x_s-t)^2 +r_s^2]^{5/2}}  \dd t  \\
     & \qquad \qquad \quad + \frac{ -y_s^2 (s-x_s) } {[(x_s-s)^2+r_s^2]^{5/2}} \int_{-\Lambda/2}^{\Lambda/2}  \frac{ 6 (x_s-t) r_s^2} {[(x_s-t)^2 +r_s^2]^{5/2}}  \dd t\\
    & \qquad \qquad \quad + \frac{- y_s^2 (s-x_s)^2} {[(x_s-s)^2+r_s^2]^{5/2}} \int_{-\Lambda/2}^{\Lambda/2}  \left. \frac{-  3 (x_s-t)^2} {[(x_s-t)^2 +r_s^2]^{5/2}} \dd t \right\} .
\end{split}
\end{equation}

We note that
\begin{equation}
    J_{0,3} = \int_{-1/2}^{1/2} \frac{1} {[(x_s-t)^2 +r_s^2]^{3/2}} \dd t = \frac{1+2z}{r^2[(1+2 z)^2+4 r^2]^{1/2}}+ \frac{1-2z}{r^2[(1-2 z)^2+4 r^2]^{1/2}} . 
\end{equation}

Averaging over the length of the cylinder, ($ -\Lambda/2 < s < \Lambda/2$), and using $L$ as unit length in the integrals, we get
\begin{equation}
\begin{split}
    u^{(s)}_{\perp,y}  (\mathbf{x}_s) & =   - \frac{ 5 \xi_{\perp}^{(0)} U d^3 }{16 \pi \mu \Lambda^3 }   y_s^2 \left[- J_{0,3}^2 +3 J_{0,5}^2 r_s^4 +6 J_{1,5}^2 r_s^2 +3 J_{2,5}^2 \right] \\
    & = - \frac{ 5 \xi_{\perp}^{(0)} U d^3 }{16 \pi \mu \Lambda^3 }  \frac{4 y_s^2}{3 r_s^4} \left[ 1 + \frac{1 +12 (r_s^2-x_s^2) + 48  (x_s^4+r_s^4) -64(r_s^2+x_s^2)^3}{\left[1+8  \left(r_s^2-x_s^2\right)+16 \left(r_s^2+x_s^2\right)^2\right]^{3/2} }\right], 
\end{split}
\end{equation}
so the new drag coefficient is 
\begin{equation}
\begin{split}
    \xi_\perp(\mathbf{x_s}) & = \xi_\perp^{(0)} \left[ 1 + \frac{ 5 \xi_{\perp}^{(0)} d^3 }{12 \pi \mu \Lambda^3 }  \frac{ y_s^2}{r_s^4} \left( 1 + \frac{1 +12 (r_s^2-x_s^2) + 48  (x_s^4+r_s^4) -64(r_s^2+x_s^2)^3}{\left[1+8  \left(r_s^2-x_s^2\right)+16 \left(r_s^2+x_s^2\right)^2\right]^{3/2} }\right) \right] \\
    & =  \xi_{\perp}^{(0)} \left( 1 + \delta \xi_{\perp}(\mathbf{x}_s) \right) . 
\end{split}
\end{equation}

\subsection{Suspension of many   spheres}
For a dilute suspension of non-interacting spheres at $\bm{x}_s$, $\bm{x}_1$, ..., $\bm{x}_n$, we follow classical work~\cite{batchelor1972,davis1992} and write
\begin{equation}
    \xi_{||} = \xi_{||}^{(0)} \left[ 1 + \iiint_\mathcal{V} \mathcal{P}(\mathbf{x_s}) \delta \xi_{||}(\mathbf{x}_s)  \dd \mathbf{x}_s \right],
\end{equation}
with $ \mathcal{P}(\mathbf{x_s})$ the probability of finding a sphere at $\mathbf{x}_s$. Taking uniformly distributed spheres outside the cylinder, with $r_s > a +d $ with volume fraction $\phi$, and considering only $- \Lambda/2 < x < \Lambda/2$ to neglect end effects gives
\begin{equation}
    \xi_{||}(\phi) = \xi_{||}^{(0)} \left[ 1 +  \frac{\phi}{4/3 \pi d^3} \iiint_{r_s>(a+d), \abs{x} < \Lambda/2} \delta \xi_{||}(\mathbf{x}_s)  \dd \mathbf{x}_s \right] . 
\end{equation}
In polar coordinates get the modified drag coefficient ratio 
\begin{equation}
    \xi_{||}(\phi) = 
    \xi_{||}^{(0)} \left[ 1 + \frac{5 \xi_{||}^{(0)} d^3}{12\pi \mu \Lambda^3}  \frac{3 \phi }{4 \pi d^3} 4 \pi \Lambda^3 \mathcal{I}(\frac{a+d}{L})  \right] 
    =  \xi_{||}^{(0)} \left[ 1 + \frac{15 \xi_{||}^{(0)} \phi }{12 \pi \mu}   \mathcal{I}_{||}(\frac{a+d}{L})  \right] ,
\end{equation}
where 
\begin{equation}
    \mathcal{I}_{||}(s) = \int_s^\infty \int_{s}^\infty \frac{1}{r_s} \left(1 + \frac{1 -12 (r_s^2+x_s^2) + 48 (x_s^4-r_s^4+2r_s^2x_s^2) -64(r_s^2+x_s^2)^3}{\left[1+8  \left(r_s^2-x_s^2\right)+16 \left(r_s^2+x_s^2\right)^2\right]^{3/2} }\right)  \dd r_s \dd x_s
\end{equation} 

Similarly, in the perpendicular case, 
\begin{equation}
    \xi_{\perp}(\phi) = 
    \xi_{\perp}^{(0)} \left[ 1 +  \frac{ 5 \xi_{\perp}^{(0)} d^3 }{12 \pi \mu \Lambda^3 }  \frac{3 \phi }{4 \pi d^3} 4 \Lambda^3 \mathcal{I}(\frac{a+d}{L})  \right] 
    = \xi_{\perp}^{(0)} \left[ 1 +  \frac{ 15  \xi_{\perp}^{(0)} \phi  }{12 \pi^2 \mu  }   \mathcal{I}(\frac{a+d}{L})  \right] ,
\end{equation}
with 
\footnotesize
\begin{equation}
 \mathcal{I}_{\perp}(s) = \int_{s}^\infty  \int_{0}^{2 \pi} \int_{0}^\infty \frac{1}{r_s} \left[ 1 + \frac{ 5 \xi_{\perp}^{(0)} d^3 }{12 \pi \mu \Lambda^3 }  \frac{ \cos^2 \theta_s }{r_s^2} \left( 1 + \frac{1 +12 (r_s^2-x_s^2) + 48  (x_s^4+r_s^4) -64(r_s^2+x_s^2)^3}{\left[1+8  \left(r_s^2-x_s^2\right)+16 \left(r_s^2+x_s^2\right)^2\right]^{3/2} }\right) \right]  \dd x_s \dd \theta_s \dd r_s.
\end{equation} 
\normalsize

The above expressions can be integrated analytically in the case of an infinitely long cylinder $\Lambda \rightarrow \infty$, as the change in drag coefficient no longer depends on $x_s$, by symmetry. We then also have that  $ \forall n \textrm{ odd,} \; J _{n,p} (0) = 0$.
Let us rewrite Eq.~\eqref{eqsuspar} for the sphere in the middle of the cylinder, 
\begin{equation}
\begin{split}
    u_{||,x}^{(s)} 
    = & \frac{ - 5 d^3 \xi_{||}^{(0)} U}{ 16\pi \mu L} \left[ 6 r_s^2 J_{2,5}^2  \right] \\
     = & \frac{- 5 d^3 \xi_{||}^{(0)} U \Lambda^5 }{6 \pi \mu  r_s^2 \left( \displaystyle 4 r_s^2 + \Lambda^2\right)^3}  ,
\end{split}
\end{equation}
and integrate it to obtain the modified drag coefficient, 
\begin{equation}
    \xi_{||}(\phi) = \xi_{||}^{(0)} \left[ 1 +  \frac{\phi}{4/3 \pi d^3} 2 \pi L  \int_{(a+d)}^{\infty} \frac{ 5 d^3 \xi_{||}^{(0)} U \Lambda^6  }{6 \pi \mu L  r_s \left( \displaystyle {4 r_s^2}+ {\Lambda^2}\right)^3} \dd r_s \right] .
\end{equation}
which yields
\begin{equation}
   \delta  \xi_{||}(\phi) = \frac{ 5 \phi \xi_{||}^{(0)}  }{16 \pi \mu } \left[ 2 \log \left(\frac{\Lambda^2}{4(a+d)^2}+1\right)+\frac{-8 \Lambda^2 (a+d)^2-3 \Lambda^4}{\left(4(a+d)^2+\Lambda^2\right)^2} \right]. 
\end{equation}

For the perpendicular motion, we use the same symmetries to obtain the additional speed
\begin{equation}
\begin{split}
    u^{(s)}_{\perp,y}  (\mathbf{x}_s) & =   - \frac{ 5 \xi_{\perp}^{(0)} U d^3 }{ \displaystyle 16 \pi \mu L }   y_s^2 \left[- J_{0,3}^2 +3 J_{0,5}^2 r_s^4 +3 J_{2,5}^2 \right] \\
    & =  -
    \frac{ \displaystyle 5 \xi_{\perp}^{(0)} U d^3 y_s^2 L  \left(\Lambda^4+12 \Lambda^2 r_s^2+48 r_s^4\right)}{6 \pi \mu  r_s^4 \left(\Lambda^2+4 r_s^2\right)^3} ,
\end{split}
\end{equation}
and drag coefficient modification
\begin{equation}
    \delta \xi_{\perp}(\phi)  =  \frac{5 \phi \xi_{\perp}^{(0)} }{32 \pi \mu }  \left[   2 \log \left(\frac{\Lambda^2}{4(a+d)^2}+1\right)+\frac{16 \Lambda^2 (a+d)^2+3 \Lambda^4}{\left(4(a+d)^2+\Lambda^2\right)^2}\right] .
\end{equation}

\section{Modified slender-body theory in a dilute suspension} \label{sec:appendixSBT}

\subsection{Discretized Lighthill slender-body theory}
The position of the helix and corresponding local velocity are respectively $\mathbf{x}_H(t) $ and  $\mathbf{u}_h (t)$, 
\begin{equation} \mathbf{x}_h (t) = 
\left\{ \begin{aligned}
      &R \cos(t \tan\theta /R) \\
      - & R \sin(t \tan\theta /R) \\
      & t 
 \end{aligned} \right.   \quad  \mathrm{and } \quad \mathbf{u}_h(t)  = \left\{ \begin{aligned}  -& \Omega R \sin(t \tan\theta /R)\\ - &\Omega R  \cos(t \tan\theta /R)\\  & U  \end{aligned} \right.,
\end{equation}
with $-L/2 < t < L/2$.  Here $\Omega$ is the externally imposed angular velocity,  and $U$ is the translational velocity along the helix long axis. In the first set of experiments focusing on the forces and torques on the helix, the imposed condition is $U = 0$.  In the case of a free-swimming helix,  the condition is that the sum of forces in the $z$-direction is zero, and we solve numerically for $U$.

For numerical stability, in addition to the discretization of the helix into segments of length $2 \delta$ described in the main text,  we follow Ref.~\cite{jawed2017} and use reduced forces to solve both the inverse problem $\mathbf{F} = \mathbf{A}^{-1} \mathbf{U}$. 
We assume that the forces vary smoothly along the centerline of the filament. We then solve for a reduced number $N^* = \lfloor N/3 \rfloor $ of forces, to make the inverse problem more robust, and write that 
\begin{equation}
    \mathbf{f}^p = \sum_{r=1}^{N^*} \frac{N - \abs{b r -p}}{N} \mathbf{f}^{*r}. 
\end{equation} 
with $b = N/N^*$.
More precisely, we compute a matrix $\mathbf{C}$ of size $(3N \times 3 N^*)$ so that $\mathbf{U} = \mathbf{AC} \mathbf{F}^*$ and $\mathbf{F} = \mathbf{C} \mathbf{F}^*$, and solve first for $\mathbf{F}^*$ then obtain $\mathbf{F}$.
This method ensures the stability of the numerical integration~\cite{jawed2017}.

\subsection{In a suspension}
In a suspension, we use the expression for the modified local flow given in the main text and average it 
\begin{equation}
    \mathbf{u}_{s}^q (\phi) = \frac{\phi \mathit{Vol} }{4 \pi d^3 /3 } \frac{1}{N_s} \sum_{k=1}^{N_s} \mathbf{u}_{s}^q (\mathbf{x}_s ^k) .
\end{equation}

\newpage
\section{Additional experimental data} 
\subsection{Coefficient ratio for thin helices}\label{sec:appendixthinhelix}
\begin{figure}[ht]
    \centering
    \includegraphics[width = 0.5 \columnwidth]{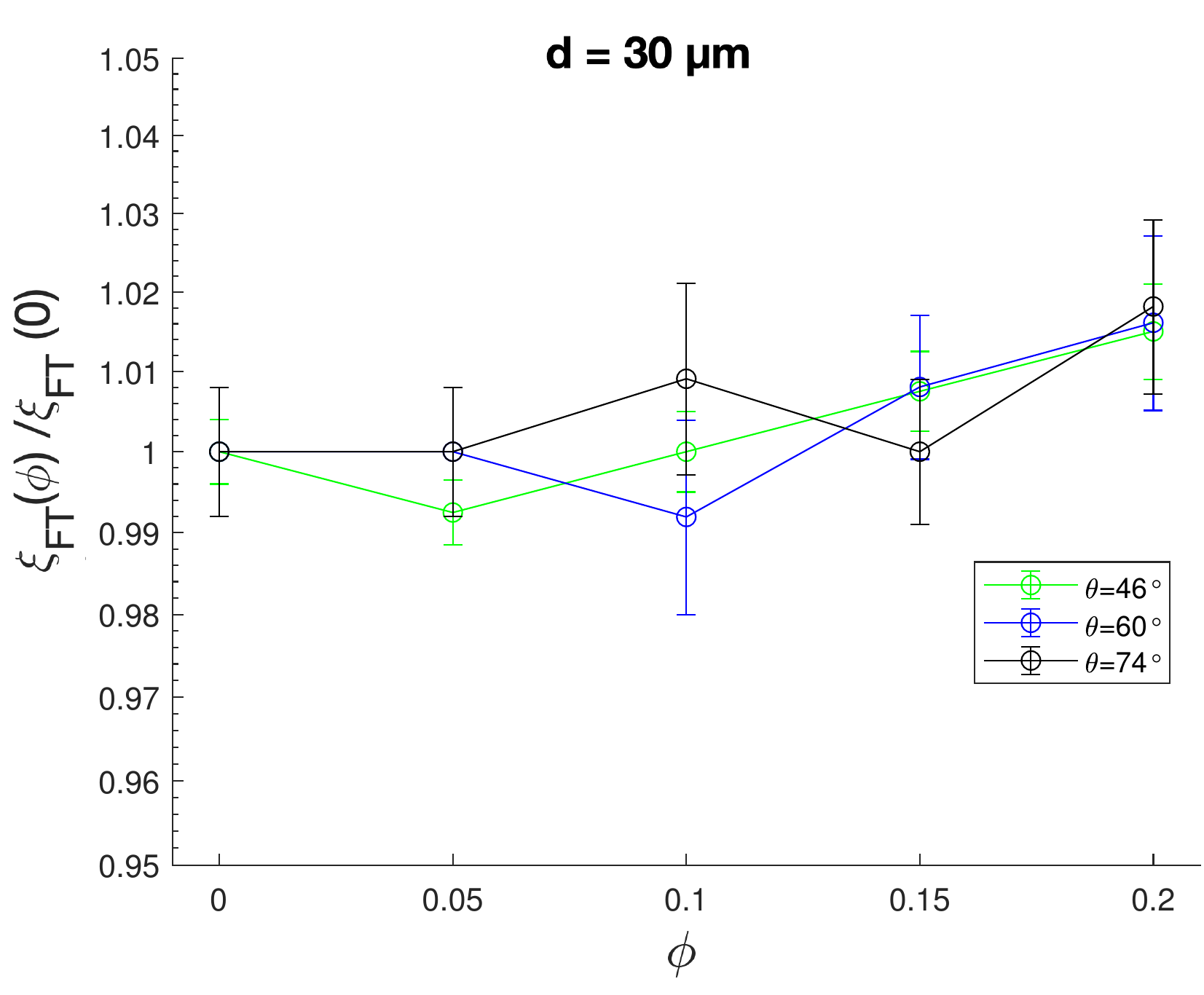}
    \caption{ Change in coefficient ratio for various helix geometries for a thinner helix with $a/R=0.06$.   
    }  
    \label{fig:app_c_coeffratio}
\end{figure}

We show in Fig.~\ref{fig:app_c_coeffratio} additional experimental data for a thin helix with $a/R = 0.06$, as opposed to $a/R=0.13$ as in the main body of the text. The thin helices are less responsive to changes in particle concentration. 
While the drag coefficient ratio $\xi_{F\Gamma}$ also tends to increase with particle volume fraction, this effect is much weaker than for the larger helices in Fig.~\ref{fig:xi_vs_phi}a. 

\subsection{Swimmer speed for all geometries}
\label{sec:appendixswimmingall}

\begin{figure}[t]
    \centering
    \includegraphics[width = .9\columnwidth]{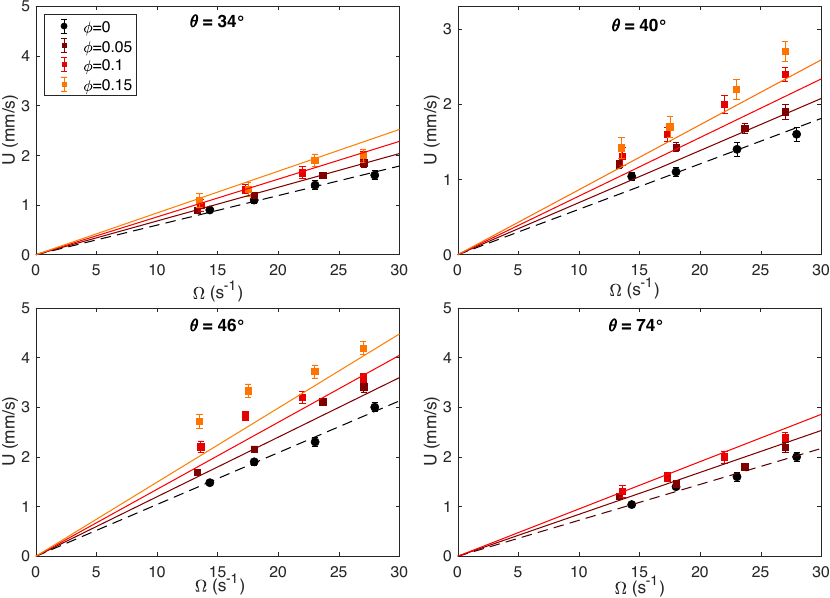}
    \caption{Raw data for the swimming speed $U$ in experiments for different helix geometries, particle volume fractions $\phi$ and angular velocities $\Omega$. The dashed lines at $\phi = 0$ show the SBT simulation with a fitted drag force from the head. The solid lines show the speed increase predictions from SBT in suspensions.}
    \label{fig:swimming_all}
\end{figure}

In addition to the measurements of swimming speed for $\theta=60^o$, experiments were conducted for swimmers with helical angles $\theta= \{34^\circ,40^\circ,46^\circ,60^\circ,74^\circ\}$. These are the same pitch angles considered for the fixed rotation helix, discussed in section~\ref{subsec:results}. These results are shown in Fig.~\ref{fig:swimming_all}, which displays the swimming speed, $U$, as a function of the angular velocity, $\Omega$. In all cases, as discussed in the main text, the particles lead to an increase in the swimming speed. The theory performs equally well for all helix angles, $\theta$.

\end{document}